\documentclass[twocolumn]{aastex631}
\usepackage{amsmath}  

\shorttitle{AASTeX v6.3.1 Sample article}
\shortauthors{Si. et al.}
\graphicspath{{./}{figures/}}

\begin{document}

\title{The transport of angular momentum for massive stars I. Formation of slowly rotating WNE stars}

\author[0009-0000-3338-0853]{Jijuan Si}
\affiliation{Yunnan Observatories, Chinese Academy of Sciences, P.O.Box 110, Kunming 650216, China}
\affiliation{University of Chinese Academy of Sciences, Beijing 100049, China}
\affiliation{Key Laboratory for Structure and Evolution of Celestial Objects, Chinese Academy of Sciences, People’s Republic of China}
\affiliation{International Centre of Supernovae, Yunnan Key Laboratory, Kunming 650216, China}

\author[0000-0002-1424-3164]{Yan Li}
\affiliation{Yunnan Observatories, Chinese Academy of Sciences, P.O.Box 110, Kunming 650216, China}
\affiliation{University of Chinese Academy of Sciences, Beijing 100049, China}
\affiliation{Key Laboratory for Structure and Evolution of Celestial Objects, Chinese Academy of Sciences, People’s Republic of China}
\affiliation{International Centre of Supernovae, Yunnan Key Laboratory, Kunming 650216, China}
\affiliation{Center for Astronomical Mega-Science, Chinese Academy of Sciences, Beijing 100012, China}

\author[0000-0001-9291-4261]{Xue-Feng Li} 
\affiliation{Yunnan Observatories, Chinese Academy of Sciences, P.O.Box 110, Kunming 650216, China}

\author[0009-0000-3338-0853]{Zhi Li}
\affiliation{Yunnan Observatories, Chinese Academy of Sciences, P.O.Box 110, Kunming 650216, China}



\begin{abstract}
The evolutionary scenario of early-type nitrogen-sequence Wolf-Rayet (WNE) stars predicts a slowly rotating subclass that typically forms after the red supergiant (RSG) phase. Their slow rotation rates are attributed to stellar winds that remove angular momentum transferred outward during core contraction. 
We incorporate improved prescriptions for internal gravity waves and the magnetic Tayler instability into single massive star evolution models. Our simulations successfully produce slowly rotating WNE stars and determine optimal parameters for both mechanisms ($A \ge 10$ for internal gravity waves (IGWs), $\alpha = 0.01$ for revised Tayler instability (TSF)). The results demonstrate that the efficiency of angular momentum transfer in massive stars is significantly enhanced compared to low-mass stars, both processes can self-consistently explain the slow rotation of WNE stars, confirming their efficiency in angular momentum redistribution and providing crucial theoretical support for the existence of this predicted stellar population.  

\end{abstract}


 
\section{Introduction} \label{sec:intro}

Rotation is one of the primary uncertainties in stellar evolution and a key physical process that alters evolutionary trajectories, as it triggers instabilities that drive internal mixing and angular momentum (AM) transport \citep{1981ApJ...243..625E, 1992A&A...265..115Z, 1998A&A...334.1000M, 2002A&A...381..923S, 2024A&A...684A.169N}.
In rotating massive stars, efficient internal angular momentum transport combined with mass loss via stellar winds typically leads to rotational spin-down \citep{1998A&A...329..551L}. This rotational evolution plays a crucial role in the formation of Wolf-Rayet (WR) stars.
Stars with initial masses higher than 20 $\mathrm{M}_{\odot}$ can potentially lose their hydrogen-rich envelopes and form WR stars \citep{2007ARA&A..45..177C, 2022A&A...664A..93D, 2024arXiv241004436S}.
The WR stars represent the evolved remnants of the massive stars that have shed their outer hydrogen layers through strong stellar winds or through Roche lobe overflow in close binary systems \citep{1986ARA&A..24..329C, 1987ARA&A..25..113A, 2005A&A...429..581M}.

The WNE star is a subtype of the WR star characterized by significant nitrogen and helium enrichment, exhibiting a surface mass fraction of nitrogen greater than that of carbon (${\text C}{\text s} < {\text N}{\text s}$, \citep{2012A&A...542A..29G}), along with surface hydrogen depletion or absence ($X_{\rm s} < 0.05$). Since nitrogen is synthesized in the core via the CNO cycle and mixed to the surface by mechanisms such as rotational mixing \citep{2000ApJ...528..368H, 2009A&A...496..841H} or convective mixing \citep{2000ApJ...528..368H}. There are two general scenarios for the formation of the WNE stars:

The first scenario involves very massive stars (typically above 80 $\mathrm{M}_{\odot}$). These stars usually evolve into WNE stars in the late main sequence (MS) phase or early core helium-burning phase \citep{2023ApJS..268...51L, 2024ApJ...974..194S}. They experience the helium core contraction and heating, while the outer envelope rapidly expands, resulting in intense mass loss. This process may directly eject the outer layers through a luminous blue variable (LBV) phase, leaving behind a rapidly rotating helium core.
The second scenario applies to stars of lower initial masses. In this case, the formation of the WNE stars occurs later, usually in the advanced stages of helium burning \citep{2003A&A...404..975M, 2012A&A...540A.144S, 2024arXiv241004436S}. This category has a high probability of following the evolutionary sequence: `O star→ Blue supergiant (BSG) (→ RSG → BSG (blue loop)) → RSG → WN star\footnote{Nitrogen-sequence WR (WN) stars. The spectral classification is based on the line ratios of N III-V and He I-II, and is divided as two types: Early-type WN stars (WNE) span WN2--WN5, while late-type (WNL) span WN7--WN9. The WN6 subtype is transitional and can be classified as either early or late  \citep{2007ARA&A..45..177C}. The WNL and WNE stars are primarily differentiated by their surface hydrogen abundances.}'\citep{2005ASPC..332....3M}.

Based on the above, we could predict the existence of two observable WNE subtypes. The two are empirically distinguished by the equivalent width of the He II $\lambda$5412 emission line: broad-lined (WNE-s) and narrow-lined (WNE-w) stars, which correspond to populations with relatively stronger and weaker surface rotation rates, respectively.
We attribute the slow rotation of the WNE-w type, which evolves from a RSG phase, to efficient angular momentum transport between the helium core and the convective envelope. During the RSG stage, the hydrogen-rich envelope is largely expelled, carrying away angular momentum transferred outward from the contracting helium core \citep{2008A&A...489..685E, 2011A&A...530A.115B, 2013ApJ...764..166D}. This process significantly spins down the core, leading to the emergence of a slowly rotating WNE star.

However, a key question remains: what is the specific mechanism that transports internal angular momentum to the outer envelope? Similarly, in low-mass stellar evolution, the observed difference between core and surface rotation suggests some unknown efficient transport mechanisms for angular momentum in the radiation zones \citep{2002A&A...381..923S, 2020A&A...634L..16D, 2022A&A...664L..16E}. Extensive research has led to three main perspectives on processes that could enhance angular momentum transport in stars:

1. Pure hydrodynamic processes \citep{2006ESASP.624E..38P, 2012A&A...544L...4E, 2013A&A...555A..54C, 2013A&A...549A..74M}. These processes operate without magnetic fields, relying solely on hydrodynamic instabilities (such as turbulent shear and large-scale circulation) to redistribute angular momentum within stellar interiors.

2. Wave-driven transport, primarily by internal gravity waves (IGWs) \citep{2014ApJ...796...17F, 2015ApJ...815L..30R, 2017ApJ...848L...1R, 2021NatAs...5..715P, 2016A&A...588A.122P, 2017A&A...605A..31P}. Mixed modes may also play a crucial role in angular momentum transport within evolved red giants \citep{2015A&A...579A..30B, 2015A&A...579A..31B, 2025A&A...699A.310B}.

3. Magnetic dynamo transport \citep{2002A&A...381..923S, 2019MNRAS.485.3661F, 2020A&A...641A.117D, 2022A&A...664L..16E, 2023A&A...677A...6M}. As noted by \citet{2024A&A...684A.169N}, the term `magnetic' here refers to a dynamo mechanism driven by magnetic instabilities.

In this work, we aim to address angular momentum transport in slowly rotating, nitrogen-rich WNE stars by separately testing two mechanisms. First, we employ additional viscosity induced by the IGWs. Second, we adopt the revised magnetic Tayler instability formalism by \citet{2019MNRAS.485.3661F} to evaluate its applicability to the massive stars.

This paper is structured as follows: Section~\ref{sec:AM} provides a general introduction to IGWs and the revised magnetic dynamo mechanism. Section~\ref{sec:model} describes the input physics. The results are presented and analyzed in Sections~\ref{sec:nu_igw} and~\ref{sec:nu_tsf}, followed by a summary in Section~\ref{subsec:sum}.


\section{Angular momentum transport by the IGW and Tayler instability} \label{sec:AM}

\subsection{IGW for angular momentum transport} \label{subsec:IGW} 

IGWs are excited at convective–radiative interfaces in stars by shear stress \citep{1981ApJ...245..286P, 1999ApJ...520..859K, 2021JFM...916A..48C}, propagating \citep{2013ApJ...772...21R} and dissipating in the radiation zone \citep{1981ApJ...245..286P}. When the IGWs propagate through radiatively stable layers, they not only induce chemical mixing \citep{2017ApJ...848L...1R, 2020ApJ...901L..18S, 2023ApJ...943..115L} but also redistribute angular momentum within stellar interiors \citep{2013ApJ...772...21R, 2014ApJ...796...17F, 2019ARA&A..57...35A}.

The role of IGWs has been extensively investigated over the past decades, particularly regarding their contribution to the angular momentum transport in low-mass stars (e.g., \citet{1997ApJ...475L.143K, 1997A&A...322..320Z, 1999ApJ...520..859K, 2005A&A...440..981T, 2005Sci...309.2189C, 2008A&A...482..597T, 2014ApJ...796...17F}). This influence also extends to the rotational evolution of giants, as evidenced by studies of young red giants near the base of the red giant branch \citep{2016A&A...588A.122P}. More recently, \citet{2023ApJ...943..115L} demonstrated that the IGWs excited at the base of the convective envelope can account for observed lithium enrichment in giants.

Massive stars typically possess a convective core, a radiative envelope, and--in certain evolutionary stages--an extended convective envelope or intermediate convective shell layers. We can reasonably infer that the IGWs can be excited at multiple convective boundaries. An open question remains: how effective are IGWs generated at various convective boundaries inside massive stars at transporting angular momentum. To elucidate the cause of the slow rotation in massive WNE-w stars, we plan to investigate the respective roles of IGWs excited by various convective zones (core and envelope).

\citet{2023ApJ...943..115L} refined the element diffusion coefficients driven by the IGWs generated at the convective envelope during the helium-burning phase. In this advancement, the empirically derived parameters for convective eddy geometry and velocity in the original formulation are replaced with explicit expressions from the $k-\omega$ model developed by \citet{2012ApJ...756...37L, 2017ApJ...841...10L}. This modification enables a more generalized representation of convective eddy dynamics across diverse stars.
The radius of an eddy is given by:

\begin{equation}
R=\sqrt{\frac{\lambda}{\rho c_{\mathrm{p}} \Omega}}, 
\label{eq:R}
\end{equation}
where $\rho$, $\Omega$, and $c_{\mathrm{p}}$ denote the density, angular velocity and heat capacity at constant pressure for cell $k$, where $\lambda$ is radiation diffusivity defined as: 
\begin{equation}
\lambda=\frac{16 \sigma T^{3}}{3 \kappa \rho},
\label{eq:lambda}
\end{equation}
here $\sigma$ is the Stefan–Boltzmann constant, $\kappa$ is the
opacity.
The revised diffusion coefficient induced by the IGWs is: 

\begin{equation} 
\nu_{\text{IGW}} = A \left( 2D_{\text{thb}} \right)^{-\frac{1}{3}} \left[ D_{\text{th}} N \left( \frac{\rho_b}{\rho} \right) \left( \frac{r_b}{r} \right)^6 \right]^2 \cdot V^{\frac{8}{3}} f^{-\frac{4}{3}},
\label{eq:Digw}
\end{equation}
where $A$ is a free numerical factor controlling the order of magnitude of $\nu_{\text{IGW}}$. \citet{2023ApJ...943..115L} adopted $A=0.1$ to explain lithium enrichment on the red giant branch (RGB). $N$ and $r$ denote the buoyancy frequency and radius of cell $k$, respectively. $V$ is velocity of eddy, which can be expressed as $R \cdot \Omega$. $f$ is a damping factor defined by Equation \ref{eq:f}.
$D_{\text{th}}$ is the thermal diffusivity defined in Equation \ref{eq:Dth}. $D_{\text{thb}}$ represents the thermal diffusivity at the bottom of the convective zone. It should be noted that the subscript `b' in the formulas denotes a convective-radiative boundary, which serves as the excitation location for IGWs. When the IGWs are excited at the convective core, the starting point is the convective core boundary. Conversely, when the IGWs propagate inward from the convective envelope, the starting point is the base of that envelope.



The formula of IGWs we employ has been validated for its reliability in studies focusing on excitation from the convective envelope during the helium-burning phase \citep{2023ApJ...943..115L}. This study aims to investigate the evolutionary process from the helium-burning phase (red supergiant stage) to the WNE star stage, a period characterized by significant angular momentum transfer. Therefore, to examine the role of IGWs during this key phase, we set the activation time of IGWs in our simulations to the core helium-burning stage. This serves a dual purpose: First, to validate whether the formula can effectively describe the impact on angular momentum transport from IGWs excited by the convective envelope during the helium-burning phase like \citep{2023ApJ...943..115L}; and second, as a preliminary test, to explore the feasibility and potential effects of extending this model to the case of IGW excitation by the convective core.
In this phase, the core has accumulated substantial angular momentum through prior contraction. By incorporating the additional viscosity parameter ($\nu_{\text{add}}$), we establish a coupling mechanism between the core and envelope, facilitating efficient angular momentum transfer from the core to the envelope. Subsequently, when strong stellar winds strip away the envelope during the RSG phase, this transported angular momentum is removed from the system.
 
To determine the operational region for the IGWs, we draw an analogy with the structure and evolution of low-mass stars during the red giant phase. We define two key boundaries for the IGW propagation region: one endpoint at the outermost layer of the convective core, and another endpoint at the bottom of the convective envelope. The propagation region of the IGWs corresponds to the radiation zone, where the Brunt–Väisälä frequency satisfies ($N^2>0$).

\begin{equation}
f = \left| \int_{r_b}^{r} D_{\mathrm{th}} N^3 \left( \frac{r_b}{r} \right)^3 dr \right|.
\label{eq:f}
\end{equation}

\begin{equation}
D_{\mathrm{th}}=\frac{16 \sigma T^{3}}{3 c_{\mathrm{p}} \kappa \rho^{2}}.
\label{eq:Dth}
\end{equation}

\subsection{Revised Tayler instability for angular momentum transport} \label{subsec:alpha}

The Spruit–Tayler (ST) dynamo \citep{2002A&A...381..923S} is a widely studied mechanism for internal angular momentum transport in differentially rotating radiative stellar envelopes. The magnetic significantly influence stellar structure and evolution by modifying heat transport, redistributing angular momentum, and enhancing chemical element mixing \citep{2001ApJ...559.1094C, 2005ApJ...629..461B, 2012ARA&A..50..107L}.
In this work, we adopt the revised magnetic Tayler instability formalism from \citet{2019MNRAS.485.3661F} (hereafter TSF). This formalism replaces the commonly used ST dynamo, which is also based on the Tayler instability but employs a different saturation mechanism. This replacement is motivated by the significantly lower energy dissipation rate in the revised prescription compared to the original one \citep{2022A&A...664L..16E}. The reduced dissipation amplifies magnetic field strengths, resulting in more efficient angular momentum transport within the revised TSF framework than in the ST dynamo.
The general expression for the viscosity coefficient due to the Tayler instability is given by equation (8) in \citet{2022A&A...664L..16E}. The revised expression from \citet{2019MNRAS.485.3661F} is as follows:
\begin{equation}
\nu_{\text{TSF}} = \alpha^3 r^2 \Omega \left( \frac{\Omega}{N_{\mathrm{eff}}} \right)^2.
\label{eq:nu_t}
\end{equation}
Here, $N_{\mathrm{eff}}$ is the effective Brunt–Väisälä frequency \citep{2002A&A...381..923S}. $\alpha$ represents a dimensionless free parameter in the revised prescription. Its value is calibrated against observational constraints from different stellar populations. For instance, \citet{2019MNRAS.485.3661F} adopted $\alpha \sim 1$ in their formalism, while \citet{2022A&A...664L..16E} found $\alpha = 6$ for subgiant and red giant. For intermediate-mass stars, a value of $\alpha \geq 2$ is required to match the observed core rotation rates \citep{2020A&A...634L..16D}. In the case of massive stars, we explore values of $\alpha$ spanning several orders of magnitude, specifically $\alpha = 0.001$, $0.01$, $0.025$, $0.5$, $1$, and $6$, to determine the optimal calibration.

Another parameter, the minimum shear $q_{\mathrm{min}}$, needs to be introduced, which is intrinsically related to $\alpha$. The $\alpha$ exerts its effect by determining the minimum shear $q_{\mathrm{min}}$. This minimum shear $q_{\mathrm{min}}$ (Equation 36 in \citet{2019MNRAS.485.3661F}) is required for the saturation of the Tayler instability. The shear parameter $q$ regulates the saturation of the Tayler instability, thereby controlling the efficiency of angular momentum transport. When the local shear exceeds this threshold ($q > q_{\mathrm{min}}$), the Tayler instability is activated, leading to efficient angular momentum redistribution that reduces both the core rotation and the shear, until the system approaches a state where $q \sim q_{\mathrm{min}}$.
\begin{equation}
q_{\mathrm{min}} \sim \alpha^{-3} \left( \frac{N_{\mathrm{eff}}}{\Omega} \right)^{5/2} \left( \frac{\eta}{r^2 \Omega} \right)^{3/4}.
\label{eq:qmin}
\end{equation}
where $\eta$ is the magnetic diffusivity.

\section{Physics of models} \label{sec:model}

In the present work, we compute theoretical stellar evolution models using the Modules for Experiments in Stellar Astrophysics (MESA) code, developed by \citet{2011ApJS..192....3P, 2013ApJS..208....4P, 2015ApJS..220...15P, 2018ApJS..234...34P, 2019ApJS..243...10P}. Our calculations utilize the \texttt{black\_hole} test suite in MESA r12115.
We simulate stars with initial masses of 25, 32, 40, 50, 60, and 70 $\mathrm{M}{\odot}$ at solar metallicity ($Z = 0.02$), all with an initial rotational velocity of ${v}/{v}_{\text {crit}}=0.4$. Each model is evolved until central carbon depletion.
Convection is treated using the standard mixing-length theory (MLT) with a mixing-length parameter $\alpha_{\text{MLT}} = 1.5$ \citep{1965ApJ...142..841H}. The Schwarzschild criterion defines the convective boundaries. Overshoot mixing beyond these boundaries is implemented using a diffusion approach with an exponentially decaying diffusion coefficient following \citet{2000A&A...360..952H}. We adopt a moderate overshoot parameter of $f_{\mathrm{ov}} = 0.016$ for all convection zones throughout all evolutionary stages.

\begin{equation}
D_{\mathrm{ov}} = D_{\mathrm{conv},0} {\rm exp} (-\frac{2z}{f_{\mathrm{ov}}H_{\mathit P, 0}} ).
\label{eq:fov}
\end{equation}
Here, $D_{\rm conv,\,0}$ represents the diffusion coefficient at the convective core boundary, $z$ is the distance to the convective core boundary, $H_{\mathit P, 0}$ denotes the local pressure scale height, and $f_{\mathrm{ov}}$ is a free parameter governing the width of overshoot regions.  

\subsection{Rotation and diffusion} \label{subsec:rot}
Rotationally induced instabilities provide an effective mechanism for the angular momentum transport. Among these, the Eddington-Sweet circulation (ES) and the secular shear instability (SSI) are dominant \citep{2012ARA&A..50..107L, 2019A&A...622A.187D}. Therefore, in our models, we consider only the contributions from the two processes. The element diffusion equation is as follows:

\begin{equation}
\frac{\partial X_{i}}{\partial t}=\frac{\partial}{\partial m}\left[\left(4 \pi \rho r^{2}\right)^{2} D_{\text {mix }}  \frac{\partial X_{i}}{\partial m}\right]+d_{i},
\end{equation}
where $X_{i}$ is the mass fraction of element $i$, and $d_{i}$ is its the generation rate. The total elemental diffusion coefficient $D_{\mathrm{mix}}$ is given by:

\begin{equation}
D_{\text {mix}} =  D_{\text {mix(non-rot)}}+f_{\text c} *D_{\text {mix(rot)}},
\label{eq:Dmix}
\end{equation}
Here, $D_{\text {mix(non-rot)}}$ represents contributions from non-rotational processes such as convection and convective overshoot. The parameter $f_{\mathrm{c}}$ characterizes the efficiency of rotational mixing and is defined as the ratio of turbulent viscosity to the diffusion coefficient for angular momentum transport via hydrodynamic instabilities \citep{2000ApJ...544.1016H, 2008ApJ...676L..29H}. We set $f_{\mathrm{c}} = 0$, thereby excluding rotational instabilities from directly influencing chemical mixing. This step aims to create an idealized condition that minimizes the feedback from chemical mixing on stellar structure, including molecular weight gradients and radii. By doing so, it ensures that models with the same initial masses exhibit a consistent radius evolution, which in turn guarantees the comparability of their angular velocity profiles.

The transport of the angular momentum within a star is treated using a diffusion approximation in MESA \citep{1978ApJ...220..279E, 1989ApJ...338..424P, 2000ApJ...544.1016H, 1983apum.conf..253Z, 2005A&A...443..643Y, 2013ApJS..208....4P, 2016ApJ...823..102C}, shown in equation \ref{eq:AM}:   
\begin{equation}
\begin{aligned}
\left(\frac{\partial \Omega}{\partial t}\right)_{m} 
&= \frac{1}{j}\left(\frac{\partial}{\partial m}\right)_{t}
\Bigg[ \left(4 \pi r^{2} \rho\right)^{2} j \bigl(\nu + \nu_{\text{add}}\bigr)
\left(\frac{\partial \Omega}{\partial m}\right)_{t} \Bigg] \\
&\quad - \frac{2 \Omega}{r} \left(\frac{\partial r}{\partial t}\right)_{m}
\left(\frac{1}{2} \frac{d \ln j}{d \ln r}\right),
\end{aligned}
\label{eq:AM}
\end{equation}
here $\Omega$ represents the angular velocity and $j$ the specific angular momentum. We introduce an additional viscosity coefficient $\nu_{\text{add}}$ into this formulation to account for contributions from either the internal gravity waves ($\nu_{\text{IGW}}$, Equation~\ref{eq:Digw}) or the revised Tayler instability ($\nu_{\text{TSF}}$, Equation~\ref{eq:nu_t}). The effects of the two mechanisms on elements mixing are not included in the current models. 
 
\subsection{Mass loss} \label{subsec:Mdot}


We employ the \texttt{Dutch} mass-loss scheme, which integrates prescriptions for both hot and cool stars to cover different temperature regimes across the Hertzsprung-Russell diagram. For models with initial masses above 50 $\mathrm{M}{\odot}$, a stellar wind mass-loss enhanced scheme due to Eddington luminosity is adopted, for more details on the mass loss basis input refer to \citet{2023ApJS..268...51L} and \citet{2024ApJ...974..194S}. According to \citet{2024ApJ...974..194S}, we set the scaling factor $\eta_{\mathrm{Dutch}} = 1.5$.
It is important to note that while observations of RSGs generally indicate high mass-loss rates, the prescription from \citet{1988A&AS...72..259D} yields rates 3 to 50 times lower than those suggested by \citet{2005A&A...438..273V} in the RSG regime \citep{2012A&A...542A..29G}. To address this discrepancy, we scale the \citet{1988A&AS...72..259D} prescription by a factor of 18.
Additionally, the mass loss enhanced due to rotation using the prescription:

\begin{equation}
\dot{M}(\Omega)=\dot{M}(0)\left(\frac{1}{1-\Omega / \Omega_{\text {\rm crit }}}\right)^{\xi}.
\end{equation}
The parameter $\dot{M}(0)$ is the mass-loss rate without rotation, and the exponent $\xi$ is set to 0.43 \citep{1998A&A...329..551L}. The critical angular velocity at the surface, $\Omega_{\text{crit}}$, is given by:

\begin{equation}
\Omega_{\text {\rm crit}}^{2}=\left(1-\frac{L}{L_{\mathrm{Edd}}}\right) \frac{G M}{R^{3}}.
\label{eq:angular velocity}
\end{equation}

This adjustment to the RSGs mass loss lowers the initial mass threshold for the WR star formation. Consequently, it helps reconcile theoretical models with the observed population of low-luminosity WR stars. The enhanced stripping of the outer envelope at lower temperatures facilitates the evolutionary transition from RSGs to pure helium stars or WNE stars.

\section{The additional viscosity: IGW} \label{sec:nu_igw}
\subsection{The IGWs Excited by the Convective Core\label{subsec:core}}

\begin{figure*}
\gridline{\fig{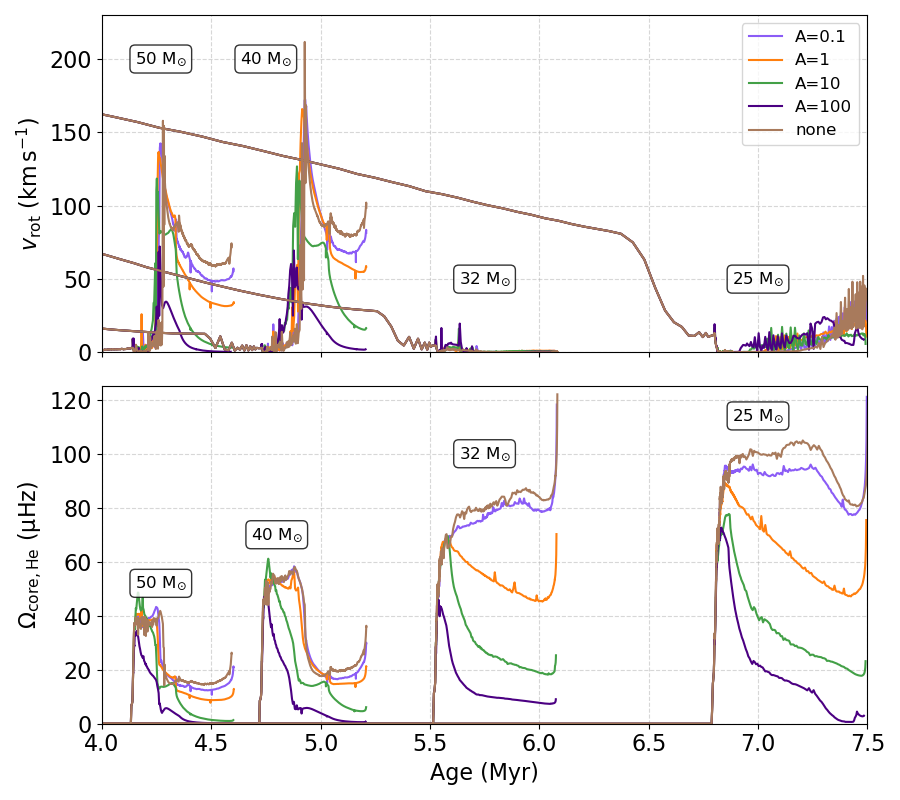}{0.49\textwidth}{(a)}
          \fig{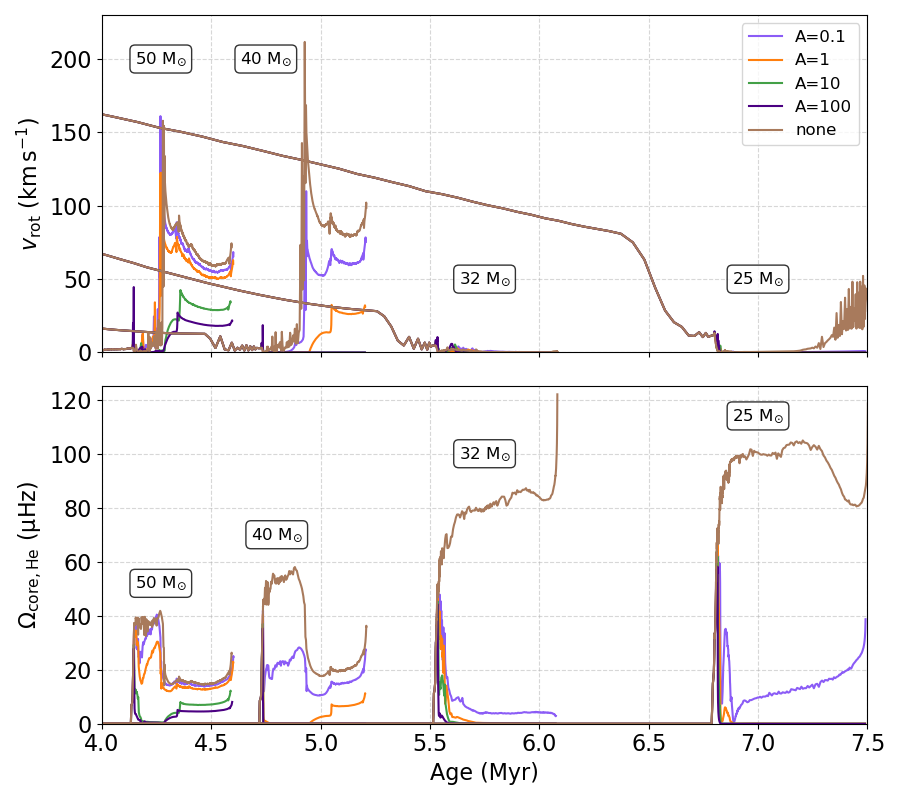}{0.49\textwidth}{(b)}    
          }
\caption{
Temporal evolution of the surface velocity (upper panel) and the helium core angular velocity (lower panel) during the core helium burning for stellar models of 25–50 $\mathrm{M}_{\odot}$. Models are arranged from longest to shortest evolutionary lifetimes. Different line styles represent varying strengths of angular momentum transport induced by the IGWs. The dark purple curve shows the baseline case without IGW-driven transport. The 25 and 32 $\mathrm{M}_{\odot}$ models do not evolve into the WNE stars, while models with initial masses $\geq$40 $\mathrm{M}_{\odot}$ can form the WNE stars. The left panel shows the results of the IGWs excited from the convective core, while the right panel shows the results of IGWs excited from the base of the convective envelope.
\label{fig:igw_v_omega}}
\end{figure*}

In the first case (hereafter Case I), we try to consider IGWs generated at the convective--radiative interface near the core, as illustrated in Figure~\ref{fig:igw_v_omega}(a). Since elemental mixing is not considered here, the stellar structure for a given initial mass remains largely consistent, with nearly identical stellar sizes throughout this phase.

We employ the free parameter $A$ from the Equation~\ref{eq:Digw} to modulate the viscosity coefficient and examine its influence on stellar rotational evolution. The results show that introducing this additional angular momentum transport mechanism effectively redistributes the core angular velocity $\Omega_\text{core, He}$ across all models, altering the angular velocity profile in the stellar core. Furthermore, for a fixed value of $A$, our results exhibit a trend consistent with the findings of \citet{2019A&A...626L...1E} and \citet{2022A&A...663A.180M} for subgiants, indicating that the efficiency of angular momentum transport varies significantly with initial stellar mass and is positively correlated with it.

For models of the same initial mass, an increase in $A$ (and thus in the diffusion strength) significantly enhances the efficiency of angular momentum transfer, resulting in a stronger suppression of the core angular velocity. In contrast, the baseline model (non-IGW model, represented by the dark purple curve and labeled `none') lacks an efficient angular momentum diffusion mechanism. Consequently, it maintains a higher core angular velocity. Toward the end of helium burning, core contraction and envelope expansion lead to an increasing stellar radius, and a sharp rise in the helium-core angular velocity ($\Omega_\text{core, He}$).

Due to the retention of the hydrogen-rich envelope in the 25 and 32 $\mathrm{M}_{\odot}$ models. Consequently, they exhibit similarly low velocity (even close to 0) during the late helium burning phase, regardless of whether additional viscosity is introduced to transfer angular momentum. However, when an additional transport mechanism via IGWs is included, the surface velocity slightly increases compared to the baseline models due to the outward transport of angular momentum from the stellar interior.
In contrast, for the 40 and 50 $\mathrm{M}_{\odot}$ models that can form the WNE stars during helium burning, the surface rotation is coupled to the core rotation, as the surface hydrogen is completely lost in the RSG stage. Consequently, as the efficiency of angular momentum transport increases, both the core and surface velocities exhibit a pronounced downward trend.

\begin{figure*} 
\centering
\includegraphics[width=0.98\textwidth]{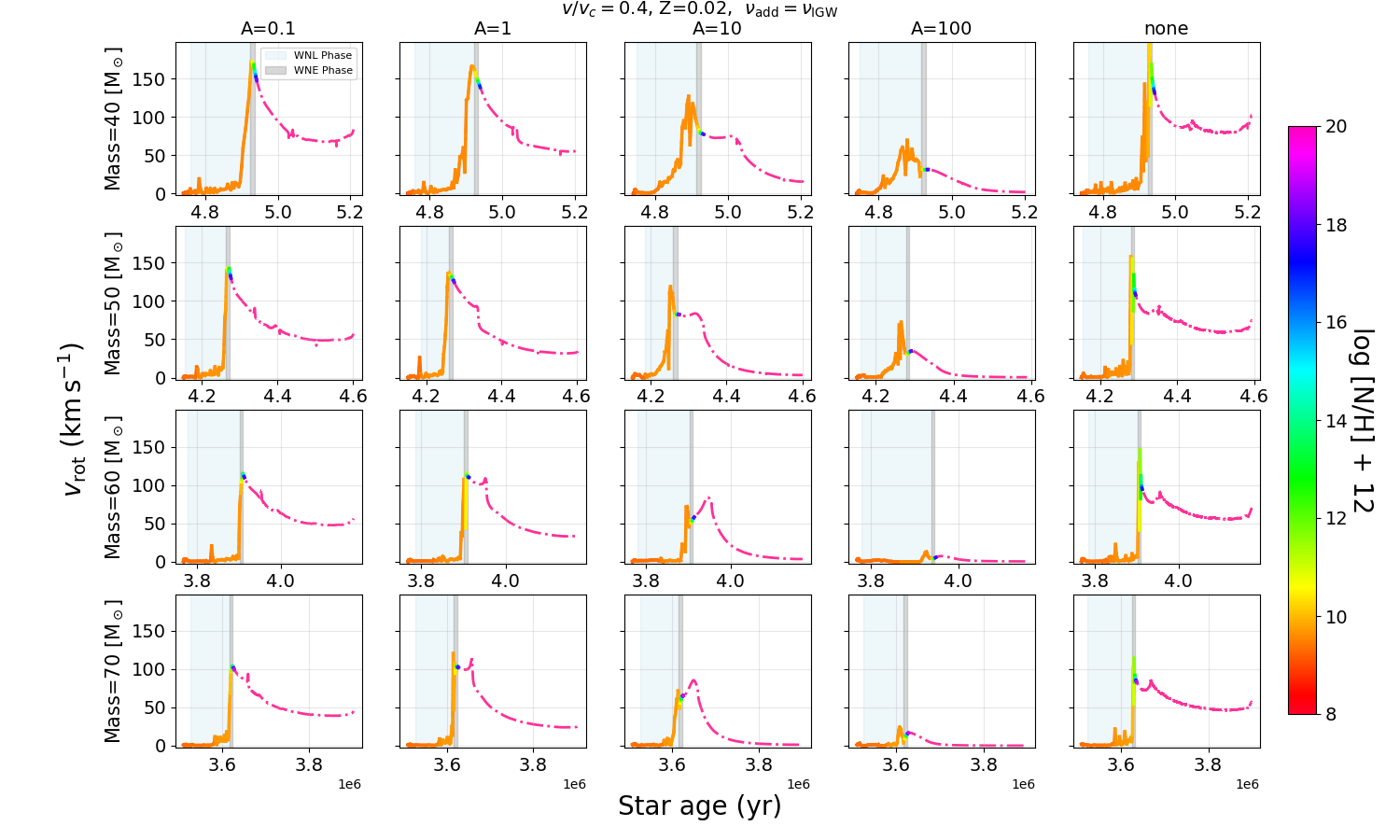}
\caption{
Evolution of surface velocity for massive ($M \geq 40 \mathrm{M}_{\odot}$) stellar models during core helium burning, for different strengths of IGWs (parameterized by A) that are excited at the convective core boundary. The color scale denotes the surface nitrogen abundance, while blue and gray shaded areas mark the WNL and WNE stages, respectively. The subsequent evolution of the hydrogen-stripped star is shown as a pink dash-dotted curve, where the surface nitrogen abundance exceeds 20 due to complete hydrogen loss. 
\label{fig:abun_N_core}} 
\end{figure*}

Based on the description of `slow rotation' in the samples of \citet{2008ApJ...676L..29H, 2009A&A...496..841H}, we adopt a reference threshold of 70 $\rm km\,s^{-1}$. We investigate stellar models with initial masses from 40 to 70 $\mathrm{M}_{\odot}$ that can form WNE stars, aiming to determine the required suitable magnitude of the parameter $A$ to slow their rotation into this observed slow regime. The results are presented in Figure~\ref{fig:abun_N_core}.

The formation of a WR star involves continuous stripping of the surface hydrogen envelope. Simultaneously, nitrogen produced by nuclear reactions is transported to the surface through further convective dredge-up and rotational mixing, leading to a significant increase in surface nitrogen abundance during the WNL\footnote{Late-type nitrogen-sequence Wolf-Rayet star, the phase preceding WNE, with $0.05 \le X_{\rm s} \le 0.4$ and $\rm{log}$ ${(T_{\text{eff}}/K)}>4.0$ \citep{2006A&A...457.1015H}} and WNE phases. By the end of the WNE phase, as indicated by the pink dash-dotted curve in figure, the nitrogen abundance has risen substantially. At this stage, the star has become a hydrogen-stripped star\footnote{All WR stars with no surface hydrogen (also called classical WR stars) are hydrogen-stripped stars, also known as helium stars (distinct from the low-mass helium stars). They are hot, helium-rich stars whose progenitors have lost their outer layers through some mechanism. However, the converse is not necessarily true, as not all stripped stars exhibit the strong winds that define WR spectral features\citep{2024arXiv241004436S}.}, and its rotational velocity continues to decrease. In this paper, the term `hydrogen-stripped' specifically refers to stars in the WNE phase and beyond, denoting helium stars without hydrogen envelope.

A pronounced spin-down occurs during the WNE phase. This is primarily because mass loss via stellar winds carries away the angular momentum transferred from the helium core to the surface, effectively braking the stellar rotation. In our baseline model without the IGWs, the WNE star would maintain a relatively faster rotation.  

\begin{figure*} 
\plotone{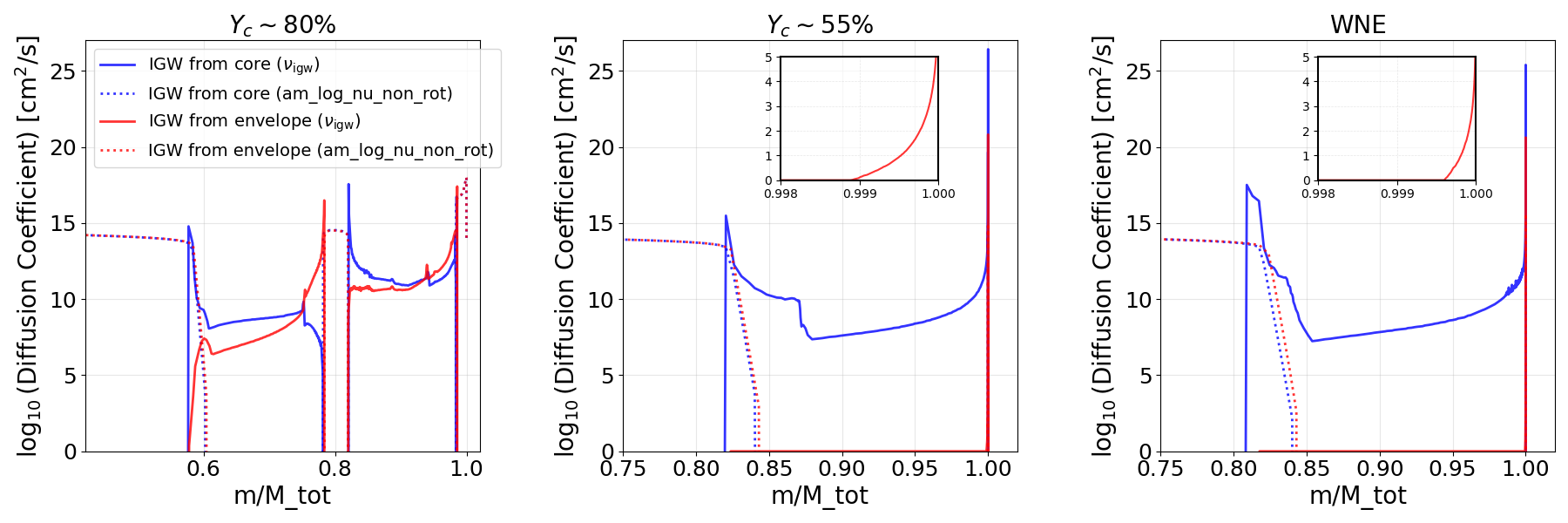} 
\caption{
Viscosity diffusion coefficient from the IGWs for the 60 $\mathrm{M}_{\odot}$ model ($A$ = 10) at different helium-burning stages. The blue and red curves represent the IGWs excited at the boundary of the convective core and the base of the convective envelope, respectively. The dotted line (labeled `am\_nu\_non\_rot') indicates the diffusion from other sources (e.g., convection-dominated), shown here as a reference baseline at the convective boundary. For the middle and right panels, the excitation of IGWs is negligible as a consequence of the convective envelope having decreased or lost.
\label{fig:nu_igw}} 
\end{figure*}

Massive stars inherently lose a significant fraction of their angular momentum through mass loss, causing them to spin down more rapidly than lower-mass stars \citep{1998A&A...329..551L, 2025arXiv250212107N}. Consequently, for more massive stars, a smaller value of the parameter $A$ is required to form a slowly rotating hydrogen-stripped star with a surface velocity below 70 $\rm km\,s^{-1}$. In other words, a less contribution from the IGW-induced diffusion is sufficient for more massive stars to achieve slow rotation. For instance, in the 40-70 $\mathrm{M}_{\odot}$ model, an $A \ge 10$ is necessary to reduce the WNE rotation speed to 70 $\rm km\,s^{-1}$. The corresponding diffusivity from the IGWs is estimated to exceed a logarithm value of 25, as shown in Figure~\ref{fig:nu_igw}.

In summary, to form a slowly rotating WNE star with a surface velocity below 70 $\rm km\,s^{-1}$, corresponding to the gray shaded region in the figure~\ref{fig:abun_N_core}, which approximates the upper limit of velocity for this population--a value of $A \geq 10$ is generally required across almost models.

\begin{figure*} 
\plotone{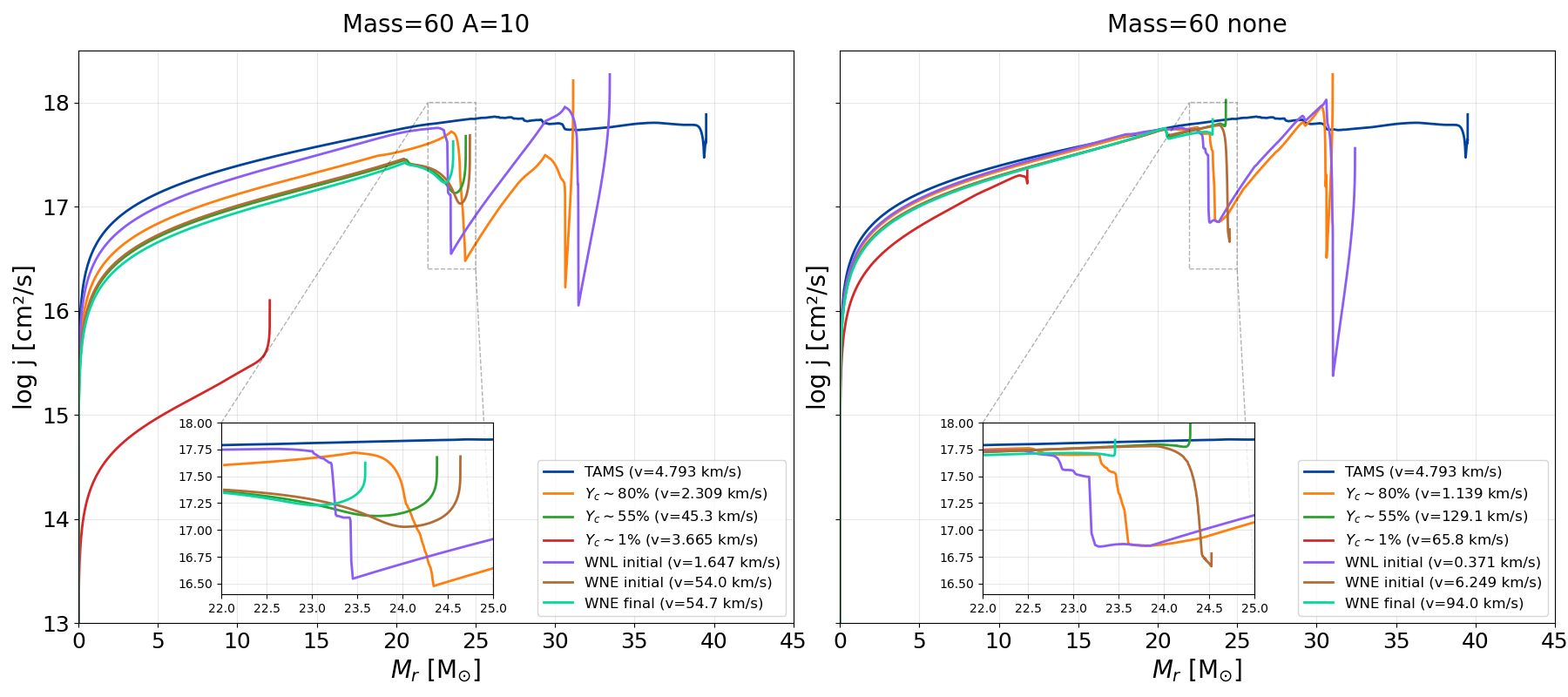} 
\caption{Specific angular momentum for the 60 $\mathrm{M}_{\odot}$ model as a function of the star mass. Solid curves show the distributions at different evolutionary stages, with labeled surface equatorial velocities. Left panel: results with the IGWs excited by and propagating outward from the convective core. Right panel: baseline case without the IGWs.
\label{fig:j_igw_core}} 
\end{figure*}

The distribution of internal specific angular momentum in the rotating 60 $\mathrm{M}_{\odot}$ model is presented in Figure~\ref{fig:j_igw_core}. We compare the angular momentum distribution influenced by the IGWs excited at the convective core with the baseline model.

In the baseline model (right panel of Figure~\ref{fig:j_igw_core}), the massive star spins down significantly after the main sequence as strong stellar winds remove most of the angular momentum. Consequently, the surface velocity drops below 5 $\rm km\,s^{-1}$ by the terminal-age main sequence (TAMS) and the early helium-burning phase (with mass fraction of center helium $Y_c \sim 80\%$). The core subsequently contracts and heats up while the envelope expands, leading to further spin-down of the envelope and spin-up of the core. By mid helium-burning ($Y_c \sim 55\%$), the core rotation reaches 129 $\rm km\,s^{-1}$. As the star evolves into a RSG and subsequently into a WNE star, enhanced mass loss further modifies the rotation velocity profile. The core velocity is maintained at 94 $\rm km\,s^{-1}$ during the WNE phase and remains at 66 $\rm km\,s^{-1}$ until helium depletion ($Y_c \sim 1\%$).

In contrast, the model with IGWs (left panel of Figure~\ref{fig:j_igw_core}, $A$ = 10) shows little difference from the baseline case at the TAMS and during the early helium-burning. However, once the IGW-driven angular momentum transport becomes active, the core velocity drops below 50 $\rm km\,s^{-1}$ by mid helium-burning ($Y_c \sim 55\%$). Throughout the WNL and WNE phases, rotation remains slower than in the baseline case, declining to below 5 $\rm km\,s^{-1}$ by helium depletion ($Y_c \sim 1\%$). The specific angular momentum decreases by nearly two orders of magnitude by the late helium-burning stage (red line), indicating that IGWs excited at the convective core efficiently transport and redistribute angular momentum within the star during core helium burning stage.

\begin{figure*} 
\centering
\includegraphics[width=0.98\textwidth]{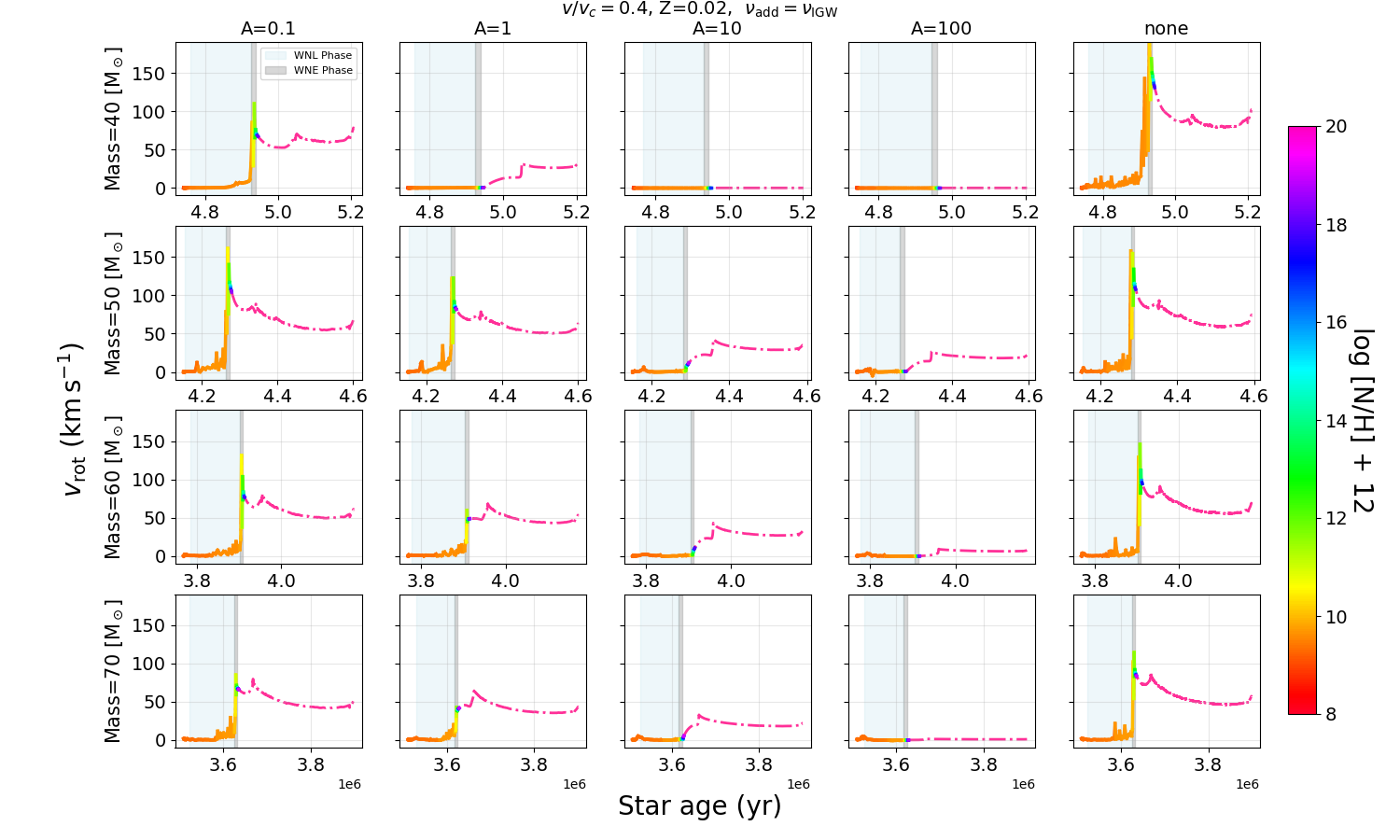}
\caption{
Similar to Figure~\ref{fig:abun_N_core}, but the IGWs excited at the base of the convective envelope.
\label{fig:abun_N_envelope}} 
\end{figure*}
 
\subsection{The IGWs Excited by the Convective envelope\label{subsec:enveloope}}
 
In Case II, we consider the IGWs excited at the base of the convective envelope. Their influence on the rotational evolution from the surface to the core is illustrated in Figure~\ref{fig:igw_v_omega}(b). We find that this configuration can also spin down the WNE or hydrogen-stripped stars. However, unlike in the Case I, the introduction of the IGWs in this scenario leads to a rapid initial decrease in the core angular velocity, followed by a gradual recovery, as seen in Figure~\ref{fig:igw_v_omega}(b).
This behavior occurs because the IGWs excited by the convective envelope do redistribute angular momentum, but the convective envelope and itself gradually diminishes over time. As the envelope recedes, the radiation zone expands and the strength of the IGW excitation weakens, as shown in Figure~\ref{fig:nu_igw}. Furthermore, the diffusion coefficient governing angular momentum transport from the envelope to the core decays significantly over the propagation distance. Consequently, the braking effect diminishes, allowing the core to gradually spin up again, as observed in Figure~\ref{fig:igw_v_omega}(b)   

\section{The additional viscosity: TSF} \label{sec:nu_tsf} 

\begin{figure*}
\gridline{\fig{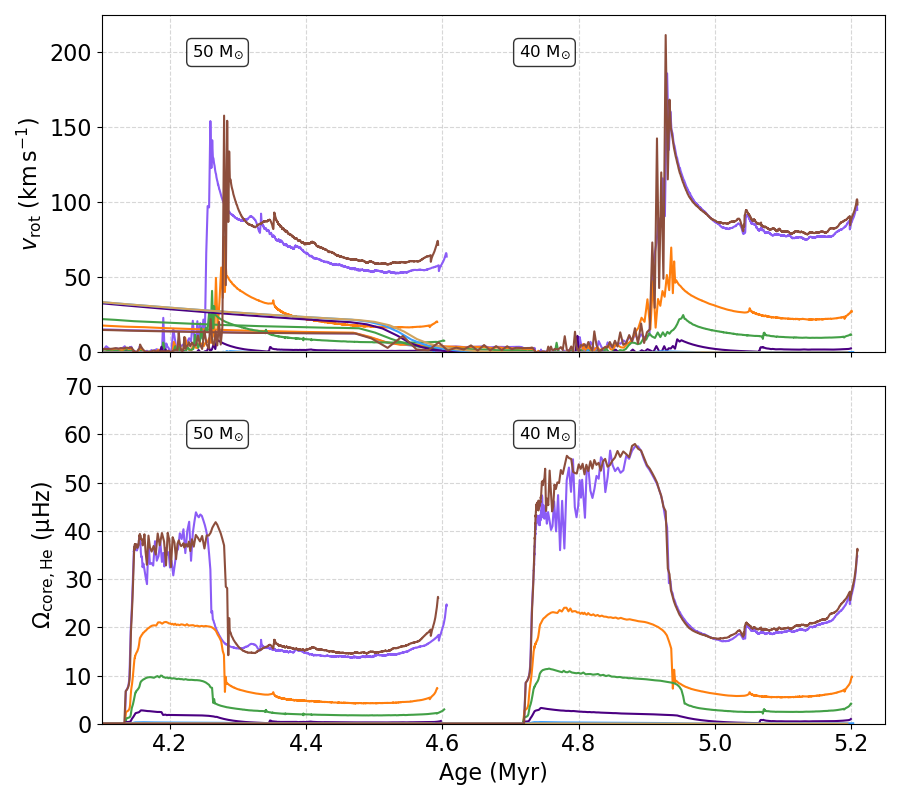}{0.49\textwidth}{(a)}
          \fig{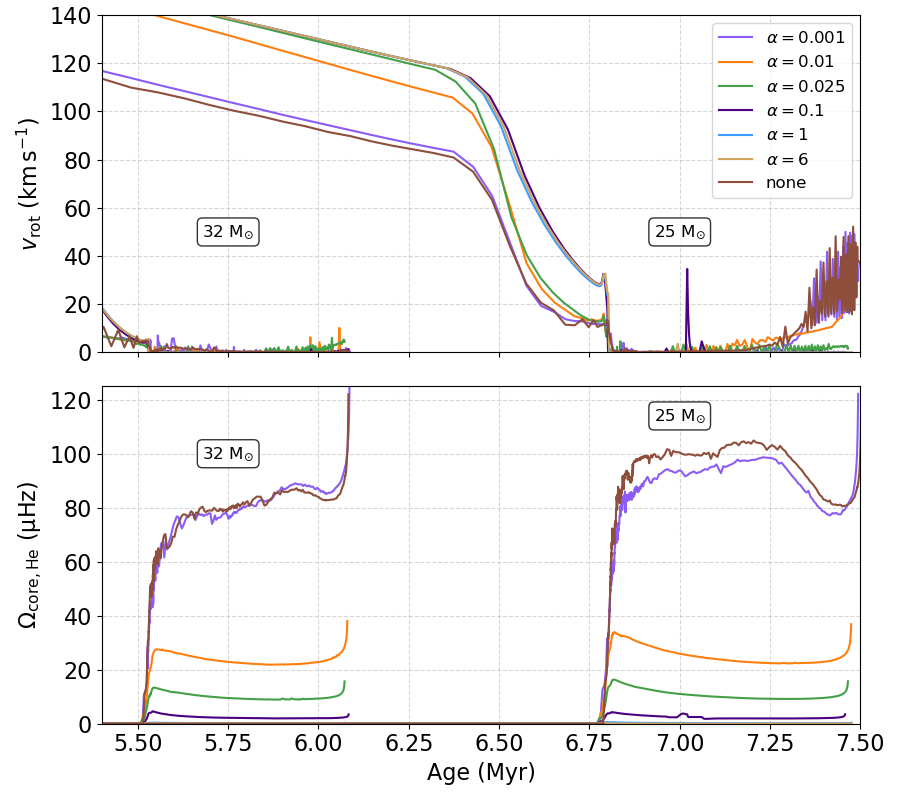}{0.49\textwidth}{(b)}    
          }
\caption{
Similar to Figure~\ref{fig:igw_v_omega}, but for 25-50 $\mathrm{M}_{\odot}$ models incorporating the revised Tayler instability for angular momentum transport. The 40, 50 and 25, 32 $\mathrm{M}_{\odot}$ models are displayed in separate panels to clarify the evolution of velocity. Note: Models with $\alpha = 1$ and $\alpha = 6$ exhibit vanishing rotation ($v_\mathrm{rot} \sim 0$, $\Omega_\mathrm{core, He} \sim 0$) throughout the evolutionary lifetime shown, rendering them indistinguishable in this figure.
\label{fig:tsf_v_omega}}
\end{figure*}

Analogous to studies of the low-mass stars, we apply the TSF formalism to the massive stars and find that it also exerts a significant braking effect. As shown in Figure~\ref{fig:tsf_v_omega}, the TSF notably influences the core angular velocity during helium burning.
For models with masses of 40 $\mathrm{M}{\odot}$ and 50 $\mathrm{M}{\odot}$, the core angular velocity increases more slowly than in the baseline model after helium-core contraction, then remains at a relatively low and stable level (around 4.2 Myr for the 50 $\mathrm{M}{\odot}$ model and 4.8 Myr for the 40 $\mathrm{M}{\odot}$ model), followed by a sharp decline, and finally maintains a lower, steady rotation rate in later evolution. In contrast, models with $M \leq 32,\mathrm{M}_{\odot}$ exhibit a monotonic and stable trend of low-speed rotation after the helium-core contraction.

Directly adopting $\alpha$ values derived from low-mass stars appears unsuitable for massive stars. When $\alpha$ is set to 1 or 6, the resulting shear $q_{\mathrm{min}}$ is very small, which allows the Tayler instability to be easily triggered and become highly efficient. This leads to an strong magnetic field amplitude that brakes the core rotation to a nearly non-rotating state. Conversely, with $\alpha = 0.001$, the value of $q_{\mathrm{min}}$ becomes much larger. Consequently, the actual shear $q$ in the stellar interior struggles to reach this high threshold, preventing the Tayler instability from being effectively excited and saturated. The resulting efficiency is therefore insufficient, and the stellar evolution resembles the baseline model.

$A$ value of $\alpha = 0.01$ is more appropriate for the massive stars during the WNE phase, producing a moderate and physically plausible spin-down evolution.
What's more, under a given value of the $\alpha$ parameter, models with different initial masses are found to exhibit similar distributions of core angular velocity during early and mid helium-burning stages. For instance, with $\alpha = 0.01$, the core angular velocity remains around 20 $\mu$Hz for 40 and 50 $\mathrm{M}_{\odot}$ models, corresponding to a surface velocity $v\leq 70$ $\rm km\,s^{-1}$ in Figure~\ref{fig:tsf_v_omega}. With $\alpha = 0.025$, the core angular velocity almost consistently remains around 10 $\mu$Hz. This seems suggests that the revised TSF model, once calibrated via the $\alpha$ parameter, provides a universally applicable prescription for the angular momentum transport across massive stars of different initial masses.
 
\begin{figure*} 
\plotone{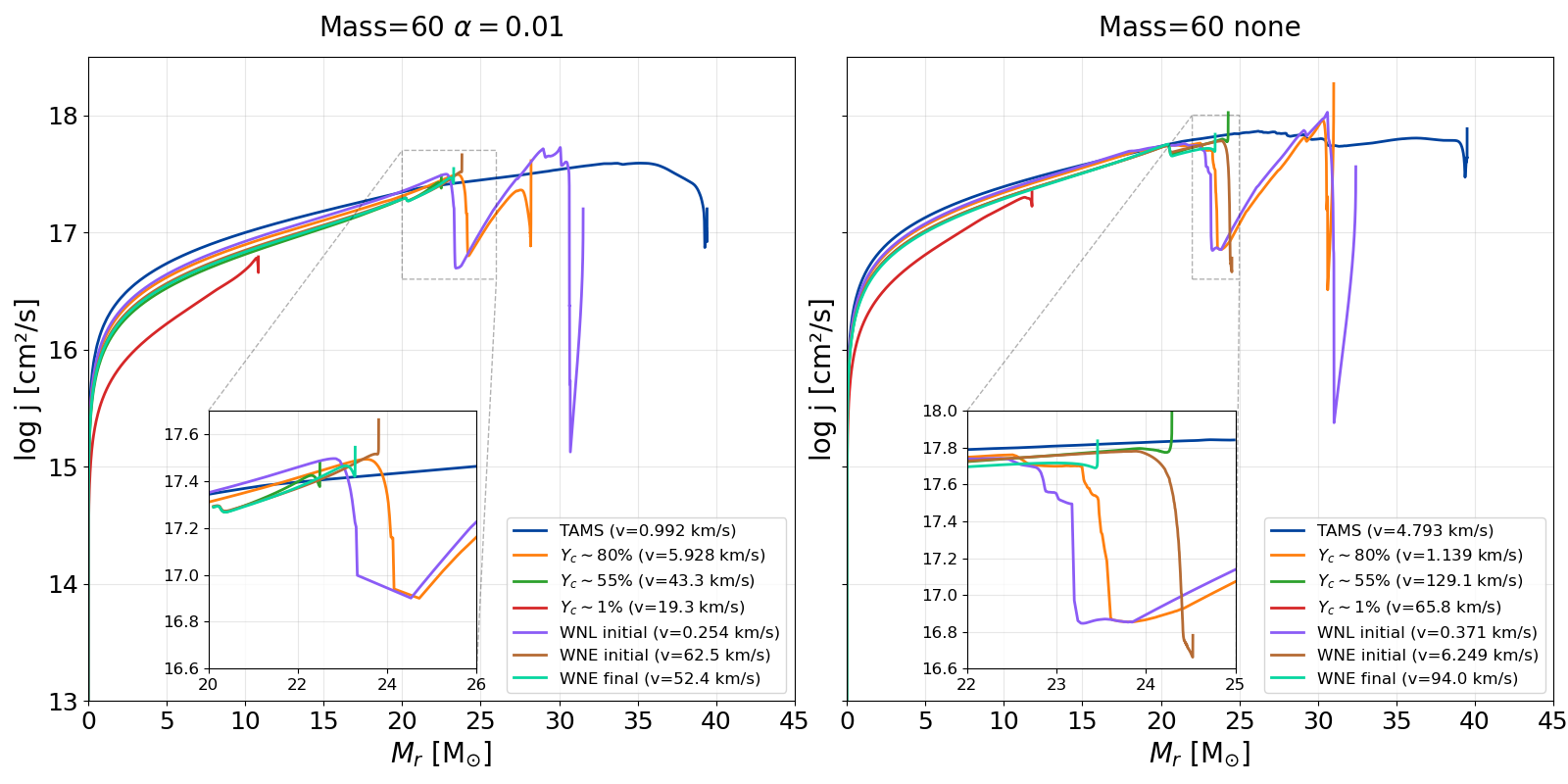} 
\caption{
Similar to Figure~\ref{fig:j_igw_core}, but for the 60 $\mathrm{M}_{\odot}$ model with the revised Tayler instability for angular momentum transport.
\label{fig:j_tsf}} 
\end{figure*}

Figure~\ref{fig:j_tsf} shows that the relationship between specific angular momentum and stellar mass, following a trend similar to that in Figure~\ref{fig:j_igw_core}. Although the velocity variation during the WNE phase is relatively modest in the TSF model, a clear spin-down is evident throughout helium burning compared to the baseline model. This demonstrates that the TSF mechanism can effectively redistribute the angular momentum during the evolution of the massive stars.

\begin{figure*} 
\centering
\includegraphics[width=0.98\textwidth]{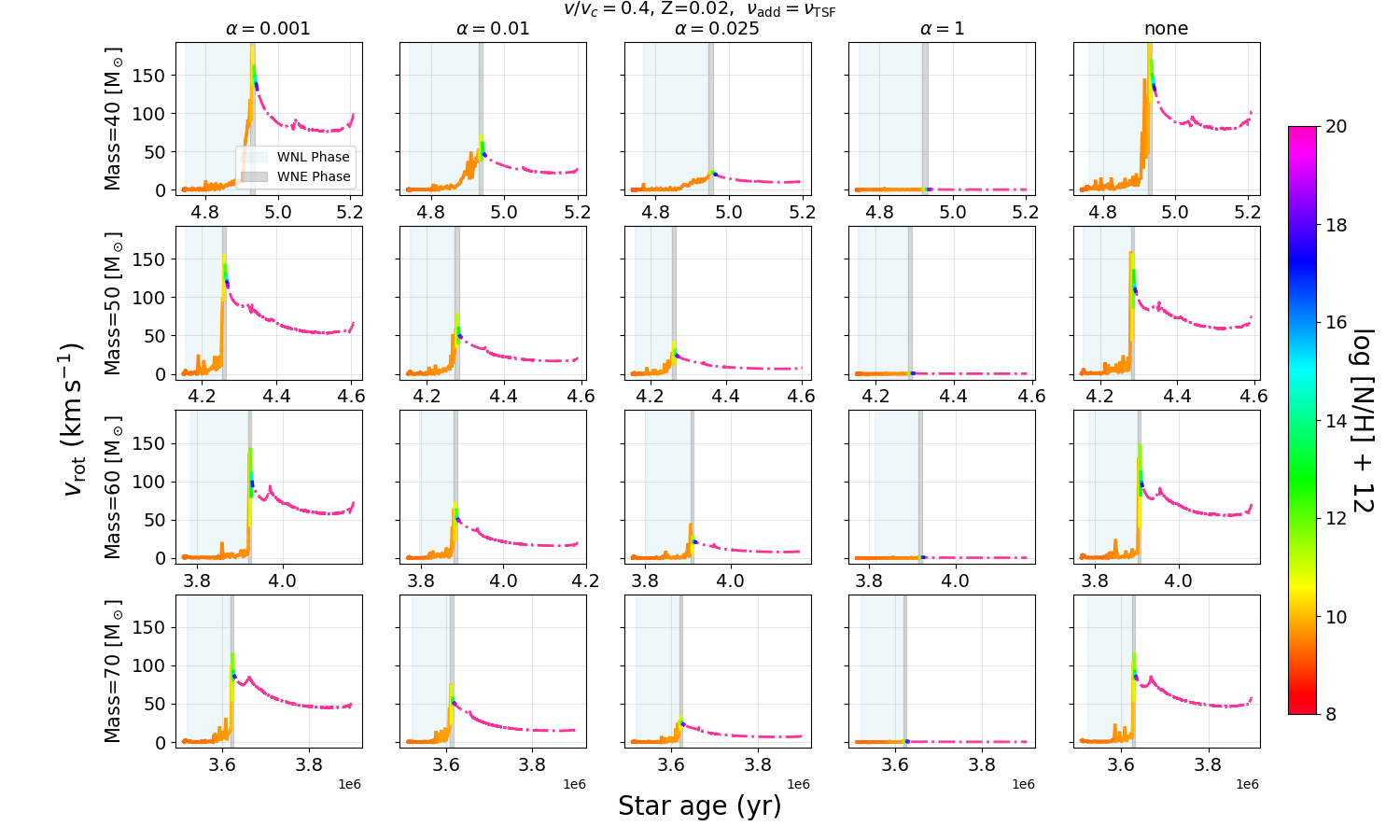}
\caption{
Similar to Figure~\ref{fig:abun_N_core}, but for 25-50 $\mathrm{M}_{\odot}$ models with the revised Tayler instability.
\label{fig:tsf_v_abun_N}} 
\end{figure*}

\section{Summary\label{subsec:sum}}

Based on the evolutionary scenario of WNE stars, this work predicts the existence of a slowly rotating counterpart formed during the helium-burning phase. This slow rotation results from the internal angular momentum transport between the core and envelope. To test this hypothesis, we simulate two main types of angular momentum transport mechanisms: the IGWs, excited at the convective core boundary and the base of the convective envelope; and the magnetic Tayler instability. We examine their influence on internal angular momentum distribution in the massive stars. Our results demonstrate that both mechanisms effectively redistribute angular momentum. Furthermore, the application to the red supergiant phase of an analogy based on the IGWs generated in the convective envelope during low-mass helium-core burning is shown to be feasible in our models. Thereby supporting the predicted existence of slowly rotating WNE stars or hydrogen-stripped stars. Besides, consistent with findings in the low-mass stars, we confirm that the transport efficiency mediated by both the IGWs and the TSF mechanism depends strongly on stellar mass.

Specifically, for a given initial stellar mass, the efficiency of angular momentum transport by the IGWs increases with the magnitude of the adopted viscosity coefficient. The free parameter $A$ in the IGWs diffusion coefficient must be at least 10 to enable the evolution of a slowly rotating WNE star. 


For the magnetic dynamo model, its transport efficiency is governed by the parameter $\alpha$. Under the criterion of 70 $\rm km\,s^{-1}$ for slow rotation, we find that, in contrast to the typical value required to explain the rotational evolution of low-mass stars ($\alpha = 1, 6$), the optimal model fit for massive WNE stars necessitates a significantly lower $\alpha$ value (the optimal value in this study is approximately 0.01).
If a calibration value close to that of low-mass stars ($\alpha = 1, 6$) is adopted, the model yields an extremely low critical shear value ($q_{\mathrm{min}}$), which substantially reduces the excitation threshold for the Tayler instability, leading to excessively efficient outward transport of core angular momentum. Given that massive stars already lose a considerable amount of angular momentum through intense stellar winds, such excessively efficient transport further exacerbates angular momentum dissipation, causing the star's rotation to nearly halt completely before evolving into the WNE stage, as seen in the $\alpha = 1$ case of Figure~\ref{fig:tsf_v_abun_N}. This contradicts the observational expectation that some WNE stars should retain limited slow rotation. Consequently, excessively large $\alpha$ values (e.g., $\alpha = 1$) are jointly excluded by the current model configuration and observational constraints. This constraint is primarily based on the observational upper limit of the surface rotation velocity of WNE stars ($\le 70$ $\rm km\,s^{-1}$). It is noteworthy that if the criterion for slow rotation or other physical inputs in the model change, the optimal $\alpha$ value may also require corresponding adjustments.

\section{Discussion}
One of the aims of this study is to attempt, from a model-prediction perspective, to assess the feasibility of the IGWs-driven angular momentum transport in forming slowly rotating WNE stars. To clarify the independent role of IGWs in the angular momentum redistribution process, we artificially neglected the chemical element mixing effects induced by them in our models. This idealized treatment, focusing solely on `pure angular momentum transport', facilitates the isolation of a single physical process. It enables us to perform controlled comparisons of the influence of different viscosity coefficients (reflecting the efficiency of transport) under similar structural evolutionary radii and timescales. The introduction of chemical mixing would cause the evolutionary trajectories of different models to diverge rapidly, rendering comparisons between mechanisms based on the same structural foundation meaningless.

However, this simplification also constitutes a major limitation of the models: Since surface abundances are central to the definition of different types of WR stars, neglecting chemical mixing may lead to deviations in the predicted surface abundances, thereby affecting indirect comparisons between evolutionary trajectories and observed states. We acknowledge that this represents a simplification of the true physical scenario to some extent.

Furthermore, applying the mathematical formalism derived from chemical diffusion in low-mass stars to describe IGWs-driven angular momentum transport in massive stars can be regarded as a modeling attempt based on structural analogy during the helium-burning phase. While the two share similarities in the form of the diffusion equation, directly transplanting this formalism and focusing solely on angular momentum transport requires more complete, self-consistent theoretical support. Future work is necessary to self-consistently treat the coupled transport of chemical abundances and angular momentum by IGWs within a unified framework, in order to more accurately reveal their full role in the late evolutionary stages of massive stars, particularly during the formation of WR stars.

Another primary limitation of this study stems from the lack of a sufficiently large sample of WNE stars with reliable projected rotational velocity measurements, owing to the evolutionary characteristics and observational properties of massive stars. To address this, we adopt the 70 $\rm km\,s^{-1}$ threshold, which is based on observations of B-type stars by \citet{2009A&A...496..841H}. This choice is grounded in the following evolutionary connection: the slowly rotating, nitrogen-enriched B-type stars from \citet{2009A&A...496..841H} are considered products of intense internal angular momentum transport (e.g., via magnetic \citep{2005ApJ...626..350H, 2008ApJ...676L..29H, 2011A&A...525L..11M}, shear \citep{2003A&A...404..975M} or binary \citep{2008IAUS..250..167L}) during the red supergiant phase \citep{2011A&A...530A.115B, 2012ARA&A..50..107L}. Although these B-type stars may still retain hydrogen-rich envelopes (as inferred from surface gravity), and thus differ in surface conditions from hydrogen-poor WNE stars, evolutionary models suggest they may share a common evolutionary pathway as `post-red supergiants' \citep{1995A&A...293..427C, 2011A&A...530A.115B, 2012ARA&A..50..107L}.

Based on this association, we propose a reasonable evolutionary hypothesis: if a stellar model can form a slow-rotating ($\le 70$ $\rm km\,s^{-1}$), nitrogen-enriched progenitor star at the B-type stage, then, as it evolves further along the `post-red supergiant' channel and undergoes intense mass loss (which eventually strips the hydrogen envelope) and additional angular momentum dissipation, it is highly likely to become an even slower-rotating WNE star. Therefore, the 70 km/s threshold does not directly define the rotational velocity of WNE stars, rather, it serves as an `estimated criterion' applied at the evolutionary starting point to identify those combinations of model parameters ($A$ and $\alpha$) capable of successfully evolving into such slow-rotating WNE stars. Consequently, this threshold provides a relevant constraint for modeling the single-star formation channel of WNE stars that originate from the post-red supergiant.
Nevertheless, it must be emphasized that this threshold remains a ‘reference value’, as direct observational verification of WNE star rotational velocities is currently lacking. Future high-quality spectroscopic observations of WR stars, particularly reliable measurements of their rotation, are essential for further testing and constraining angular momentum transport models of this kind.

Beyond the points mentioned above, future research should also address additional limitations of this study. A deeper understanding and calibration of the underlying physical mechanisms of the model are needed, with the ultimate goal of establishing a unified physical framework capable of reasonably explaining the angular momentum transport problem across the entire evolutionary stages of massive stars.

%



\section*{Acknowledgments}






This work is supported by the B-type Strategic Priority Program of the Chinese Academy of Sciences (Grant No. XDB1160202), the National Natural Science Foundation of China (Grant No. 12288102), and by the National Key R\&D Program of China (Grant No. 2021YFA1600400/2021YFA1600402). The authors also gratefully acknowledge the supports of NSFC of China (Grant Nos. 12133011 and 12273104), and the China Manned Space Program (grant no. CMS-CSST-2025-A14/CMS-CSST-2021-B06), Yunnan Fundamental Research Projects (Grant No. 202401AS070045), the International Centre of Supernovae, Yunnan Key Laboratory (No. 202302AN360001).

X.-F. L. acknowledges support from NSFC grant No. 12503043.
Si is grateful to Prof. Han-Feng Song and Dr. Facundo D. Moyano for instructive discussions regarding stellar rotation.

    $Software$: MESA version 12115 \citep{2011ApJS..192....3P, 2013ApJS..208....4P, 2015ApJS..220...15P, 2018ApJS..234...34P, 2019ApJS..243...10P}.







\bibliography{sample631}{}
\bibliographystyle{aasjournal}



\end{document}